\documentclass[a4paper,fleqn,usenatbib]{mnras}

\usepackage{newtxtext,newtxmath}

\usepackage[T1]{fontenc}
\usepackage{ae,aecompl}

\usepackage{graphicx}	
\newcommand{\xten}[1]{\mbox{$\times 10^{#1}$}}

\newcommand{\ltappeq}{\raisebox{-0.6ex}{$\,\stackrel
{\raisebox{-.2ex}{$\textstyle <$}}{\sim}\,$}}
\newcommand{\gtappeq}{\raisebox{-0.6ex}{$\,\stackrel
{\raisebox{-.2ex}{$\textstyle >$}}{\sim}\,$}}

\title[Water ice deposition and growth]
{Water Ice Deposition and Growth in Molecular Clouds}

\author[J. M. C. Rawlings and D. A. Williams]
{Jonathan M. C. Rawlings,$^{1}$\thanks{E-mail: jcr@star.ucl.ac.uk}
and D. A. Williams,$^{1}$\\
\\
$^{1}$Department of Physics and Astronomy, University College London,
Gower Street, London, WC1E 6BT, UK}

\date{Accepted 2020 November 9. Received 2020 November 9; in original form 
2020 August 10}

\pubyear{2020}

\begin{document}
\label{firstpage}
\pagerange{\pageref{firstpage}--\pageref{lastpage}}
\maketitle

\begin{abstract}
In interstellar clouds the deposition of water ice onto grains only occurs at visual 
extinctions above some threshold value ($A_{\rm th}$). At extinctions greater than 
$A_{\rm th}$ there is a (near-linear) correlation between the inferred column density 
of the water ice and $A_{\rm V}$.
For individual cloud complexes such as Taurus, Serpens and $\rho$-Ophiuchi, 
$A_{\rm th}$ and the gradients of the correlation are very similar along all lines 
of sight.
We have investigated the origin of this phenomenon, with careful consideration of 
the various possible mechanisms that may be involved and have applied a full chemical 
model to analyse the behaviours and sensitivities in quiescent molecular clouds.
Our key results are: (i) the ubiquity of the phenomenon points to a common cause, so 
that the lines of sight probe regions with similar, advanced, chemical and dynamical 
evolution, (ii) for Taurus and Serpens; $A_{\rm th}$ and the slope of the correlation 
can be explained as resulting from the balance of freeze-out of oxygen atoms and 
photodesorption of H$_2$O molecules. No other mechanism can satisfactorily explain 
the phenomenon, 
(iii) $A_{\rm th}$ depends on the local density, suggesting that there is a correlation 
between local volume density and column density,
(iv) the different values of $A_{\rm th}$ for Taurus and Serpens are probably due to 
variations in the local mean radiation field strength, (v) most ice is accreted onto 
grains that are initially very small ($<0.01\mu$m), and (vi) the very high value of 
$A_{\rm th}$ observed in $\rho$-Ophiuchi cannot be explained in the same way, unless 
there is complex microstructure and/or a modification to the extinction characteristics. 
\end{abstract}


\section{Introduction}
\label{sec:intro}

Interstellar water ice was first detected in 1973 in the infrared spectra of protostars 
by absorption in the O-H stretch mode near 3$\mu$m \citep{GF73}. Water ice had been 
sought without success in low-density diffuse regions of interstellar space, but is 
now well-known to be very abundant in dense clouds and star-forming regions where it 
contains a large fraction of the available oxygen that is not already locked in CO 
\citep[see the review by][and references therein]{vD13}. Grains of pure ice do not 
exist; rather, water ice is located as mantles on dust grains of carbon or silicates. 
It is clear that water ice is formed and retained through reactions on the surfaces of 
dust grains, rather than the deposition of H$_2$O molecules from the gas phase 
\citep[e.g.][]{JW84,Cup10,Lam13}. While water is observed to be the dominant component 
of interstellar ice, other significant component species include CO (deposited directly 
from the gas phase) and CO$_2$ (formed in surface reactions). However, water ice is 
apparently the first solid material to be deposited on grains.

For any particular interstellar cloud, the observationally determined abundance of the 
water ice shows two 
remarkably consistent characteristics as determined from the correlation between 
the 3$\mu$m optical depth ($\tau_{3\mu m}$) and interstellar extinction ($A_{\rm V}$). 

Firstly,
it appears that water ice begins to be deposited on dust grains at a critical threshold 
visual extinction, $A_{\rm V}=A_{\rm th}$ magnitudes, within that cloud \citep{Wh13}. 
Although there is 
some scatter in the vicinity of $A_{\rm th}$, this value is well-defined, so that
although $A_{\rm th}$ varies from cloud to cloud, cores within each cloud complex have 
the same value of $A_{\rm th}$ with very few outliers.
Below this cut-off value, $\tau_{3\mu m}=0$; i.e. H$_2$O ice is not present.
Ice deposition in the quiescent dark cloud in Taurus has been extensively studied
\citep{Wh89,Wh01,Wh07,Chiar94,Bergin05}
and the critical value of visual extinction for ice deposition in that cloud is 
3.2$\pm 0.1$ mag. A study of the quiescent dark cloud complex IC 5146 finds the same 
critical value \citep{Chiar11}, implying a common mechanism in regions that do not 
possess internal heating/radiation sources.
However, on other lines of sight, particularly those in star forming regions, the 
critical value may be significantly larger; for example, in the active star-forming 
Serpens cloud the critical value of visual extinction is $\sim$ 6 mag while in the 
intermediate/high-mass star-forming region of $\rho$-Ophiuchus it is $\sim$13 mag 
\citep{WB80,EH89,Tan90}.

Threshold extinctions have also been determined for other ice components, such as 
CO and CO$_2$, as well as for complex organic molecules (COMs), although these
are typically more variable and larger (especially for the COMs) as compared to 
that for water ices. 

Secondly, on paths with visual extinctions greater than the critical value,
the quantity of ice as measured by the optical depth at 3 $\mu$m rises almost linearly 
with $A_{\rm V}$ \citep{Wh89},\citep[see also][for a summary]{Wh03}, and 
the value of the slope of the plot of $\tau$(3 $\mu$m) vs. $A_{\rm V}$, which we 
denote by $\alpha_{\rm H_2O}$, appears to be very similar in each case.
Thus, $\alpha_{\rm H_2O} = 0.072 \pm 0.002$ and $0.068 \pm 0.003$ for IC\,5146 and 
Taurus, respectively \citep{Chiar11,Wh01}, although \citet{Chiar07,Chiar11} also note 
that the slope is generally shallower in dense clouds than it is in more diffuse clouds.
The similarity of the slopes might imply that all available oxygen has been converted 
to water ice. However, as we discuss below, differences in astrophysical parameters 
such as local density and local metallicity would be expected to have significant
effects. 

While ice deposition has received much theoretical attention 
\citep[e.g.][]{CH07,Ietal08,And08,Du12,Kal15}, 
the actual processes controlling the onset of ice deposition and the removal of H$_2$O 
molecules from ice remain matters for discussion. 

A number of studies have addressed the issue of the $A_{\rm th}$ phenomenon
and have usually invoked the obvious $A_{\rm V}$-dependence of photodesorption of
ices as the most likely cause, although there has been little attempt to 
explain (a) its value and universality within molecular clouds, and (b) 
why it varies between clouds. In addition, there are other possible causes 
for the phenomenon that are not directly related to the photodesorption effect 
that should also be considered. 

In a study of the effects of selective desorption mechanisms on the 
growth and composition of interstellar ices \citet{Kal15} developed
a comprehensive model of the processes on and in (the top four layers) of
ices including the full range of desorption processes that we describe in 
this paper, surface chemistry, diffusion and the chemistry in sublayers
together with considerations of the mobility of species on the surface and 
within the ices.
The model examined the time-dependencies of the ice abundances (the main components, 
and also the complex organics), in the 
context of a simple isothermal free-fall collapse model, so that - in this model - 
the extinction is directly related to the dynamical age, although a complicating factor 
is introduced here in that, as the collapse proceeds, the dynamical 
timescale becomes significantly shorter than the chemical timescale.
The paper showed  the relative importances of photodesorption and reactive desorption 
processes in determing the efficiency of freeze-out and the composition of the ices.
A value of $A_{\rm th}\sim 8-10.5$ is determined for CO ices, although the work
does not specifically address the issue of the determination and variation of
the critical extinction for H$_2$O ices.

A number of other mechanisms have recently been proposed concerning the deposition 
of interstellar ice, in addition to studies of the type discussed in the papers listed 
above. In this paper, we wish to examine the various proposals, with a view to 
addressing the following questions:

\begin{enumerate}  
\item What are the physical conditions that determine the critical extinction for ice 
deposition, and why is $A_{\rm th}$ the same along different sight lines in the
same cloud complex?
\item Why is there a wide range of $A_{\rm th}$ in different clouds?
\item Can a single mechanism account for the observed wide range of $A_{\rm th}$ in 
different clouds?
\item Can any of the suggested mechanisms limiting ice deposition be discounted?
\item What determines the slope of the $\tau$(3$\mu$m) v. $A_{\rm V}$ plot 
($\alpha_{\rm H_2O}$) and why is a single correlation apparently applicable to 
many sources, with little scatter?
\item Can we use information contained in the value of $A_{\rm th}$ and 
$\alpha_{\rm H_2O}$ to determine the physical conditions where ice deposition 
is occurring?
\end{enumerate}

As the empirical basis for testing our hypotheses and models we use the data 
summarised in \citet{Wh13}. 
They argue that the optical depth in the 9.7$\mu$m silicate absorption 
feature is the best proxy for the dust column density and deduce a value of
$A_{\rm th}=2.7\pm 0.5$ for H$_2$O for the combined datasets for L183, Taurus and
IC5146. 
In addition, the correlation between $N$(H$_2$O) and $\tau_{9.7\mu m}$ is linear and 
remarkably similar for the three sources, and from Figure 14 of that paper we can 
therefore also deduce that
\begin{equation}
N(H_2O) \sim 2.05\xten{17}\left( A_{\rm V} - 2.7 \right)~{\rm cm}^{-2}
\label{eq:column}
\end{equation}
This correlation has remarkably little scatter, and there
is little variation of either $A_{\rm th}$ or $\alpha_{\rm H_2O}$
between the three sources that are compared in that paper (the quiescent, chemically 
evolved pre-stellar core L183 (L134N), Taurus and IC\,5146).

In the discussion that follows we must recognise that the observed (i.e. line of 
sight) extinction to a source is not necessarily the same as the `local' 
extinction, that is to say the extinction of the source relative to the interstellar 
radiation field, so there will always be some uncertainty/natural scatter in the 
values of $A_{\rm th}$.
However, the near-universality of these dependencies implies that the relevant 
processes must be largely independent of the dust properties; such as the size 
distribution and the dust grain morphology and porosity etc. 

In the following Section we describe the nature of interstellar ices and the 
criterion for ice mantle growth. In Section~\ref{sec:mechanisms} we describe the 
various microscopic mechanisms that determine $A_{\rm th}$, and in 
Section~\ref{sec:analytical} we compare the relative importances of these 
processes. Section~\ref{sec:model} describes our chemical model, and the results 
are presented in Section~\ref{sec:results}. Possible alternative causes of the 
$A_{\rm th}$ phenomenon are discussed in Section~\ref{sec:alt}. Our conclusions 
and discussion are presented in Section~\ref{sec:conc}. 

\section{The requirement for ice mantle growth}
\label{sec:growth}

Interstellar ices are predominantly composed of H$_2$O, CO and CO$_2$, 
with CO and CO$_2$ having abundances of $\sim 27-32$\% and $\sim 19$\% 
respectively, relative to H$_2$O in Taurus and low-mass YSOs 
\citep[][Table 4]{Chiar11}. CO, being more volatile, is typically less abundant in high 
mass YSOs.
Other species (such as CH$_3$OH, NH$_3$, OCN$^-$, and other organics/complex organic 
molecules) typically account for $\ltappeq$10\% of the ice.

The relative abundances of the major ice components are fairly similar in different 
environments, although real differences exist and there are often significant variations 
in the abundances of the complex organics.
Whilst there is evidence for significant source to source variations in the
relative abundances of H$_2$O, CO and CO$_2$, the total contribution of
O-bearing ice molecules to the oxygen budget follows similar trends in different 
souces \citep[e.g.][]{Wh09}. 

The chemistry of the main components is straightforward;
CO ice is believed to be formed (primarily) due to the freeze-out
of CO from the gas phase. By contrast, H$_2$O and CO$_2$ are probably 
formed by surface reactions. At least four distinct H$_2$O formation mechanisms are 
possible \citep{vD13} but, in the relatively low density/high hydrogen atom abundance 
environments that we are considering in this study, the main channel is simple 
hydrogenation:
\[ {\rm O + H \to OH} \]
\[ {\rm OH + H \to H_2O} \]
The latter reaction being in competition with 
\[ {\rm OH + CO \to CO_2 + H} \]

As gas-phase chemistry is not believed to be important this means that the 
formation rate of H$_2$O ice is mainly determined by the rate at which oxygen 
atoms stick to grains and are subsequently hydrogenated to H$_2$O molecules.
If these are the only relevant reactions, then it also follows that the 
abundances of H$_2$O and CO$_2$ ice should be approximately
correlated, which is indeed seen to be the case \citep[e.g.][]{Noble13}. 


Ultimately, whatever mechanism(s) is/are at play, the abundances of ices on 
grains and the explanation for the $A_{\rm th}$ phenomenon must be determined 
by the balance between the freeze-out/formation of H$_2$O and its desorption,
perhaps coupled with observational biasing and line of sight effects.

We can therefore deduce the criterion for efficient ice formation as follows:
the rate at which a species freezes-out onto dust grain surfaces obviously 
depends on the available surface area.
We make the assumptions that (i) the ice composition is dominated by H$_2$O, and 
(ii) desorption from the the ice mantle is dominated by processes that occur at 
the surface (as would be the case for photodesorption, or desorption following 
H$_2$ formation). The desorption rate then depends on the fractional composition 
of the top monolayer(s) of the ice and scales linearly with the fractional coverage 
of the surface of the grains by H$_2$O ($f_{\rm H_2O}$). 

If $f_{\rm H_2O}$ reaches unity 
(i.e. total surface coverage in a monolayer of a pure ice composition) then the 
desorption rate will saturate.
If the freeze-out and desorption rates are balanced at a level corresponding to
$f_{\rm H_2O}<1$ then, clearly, an ice mantle will not grow; the net freeze-out 
is quenched before a monolayer of ice can be formed. 

If on the other hand, when 
$f_{\rm H_2O}$ reaches 1 and the desorption rate is less than the freeze-out rate, then 
continual mantle growth will occur. Indeed, as the mantle grows the effective
surface area grows and the rate of growth will accelerate (the `snowball' effect).
The imbalance between desorption and freeze-out will be further enhanced by the fact 
that the binding energy of water molecules to ice is larger, due to strong hydrogen 
bonding, than it is for bare carbon or silicate grains \citep{WHW92}.

Thus the criterion for ice mantle growth is that the desorption rate at $f_{\rm H_2O}=1$
must be less than the freeze-out rate. If $f_{\rm H_2O}=1$ can be reached, then 
rapid ice mantle growth will ensure. 
This condition can be used to define $A_{\rm th}$.

\section{Mechanisms and processes that may determine the critical extinction}
\label{sec:mechanisms}

Various non-thermal mechanisms have been suggested as having a defining role in 
determining the critical visual extinction for the deposition of ice on dust grains. 
These include:
\begin{enumerate}
\item Non-thermal, continuous desorption,
\item The dependence of the H$_2$O-grain surface binding energy on $A_{\rm V}$,
\item The requirement for a critical O-atom flux at grain surfaces, and
\item Variations in the efficiencies of surface chemistry.
\end{enumerate}

In this study we describe these mechanisms and assess their viability as a cause of 
the $A_{\rm th}$ phenomenon.
Firstly, we need to identify and quantify the the various microscopic processes that 
may determine the abundance of H$_2$O ices. These include:
(a) gas-phase formation (and destruction),
(b) photodissociation,
(c) freeze-out,
(d) surface chemistry (forming H$_2$O, as described above, and also including 
photodissociation of the ices), and
(e) desorption.
Of these, (a) and (b) are probably relatively unimportant as explained above, whilst 
(c)-(e) depend on a number of parameters, including the local
temperature, density, radiation field strength, available dust grain surface area,
desorption yields, surface coverage of dust grains by the various ice components
etc.
It is therefore quite remarkable that a single ice band-extinction correlation exists. 
This tells us that whatever the physical origin of $A_{\rm th}$ is, it must be very 
robust to any assumptions that we make, or variabilities that exist, in the
grain physics and the chemical processes involved.

We now discuss the specific processes and assess their range of relevance/applicability,
before identifying the possible dependences on $A_{\rm V}$.

\subsection{Freeze-out}
\label{subsec:freeze}

Following \citet{RHMW}, the rate of decline (cm$^{-3}$s$^{-1}$) of the abundance of a 
gas-phase species, $i$, due to freeze-out can be written as:
\begin{equation}
\dot{n}_{i} = -\left( \frac{k}{2\pi m_{\rm H}}\right)^{\frac{1}{2}} \sigma_{\rm H}
S_{\rm i}C\left( \frac{T_{\rm gas}}{m_{\rm i}}\right)^{1/2}n_{\rm H}n_{\rm i} 
\label{eq:freeze}
\end{equation}
where: 
   $n_{\rm H}$ is the hydrogen nucleon density (cm$^{-3}$),
   $S_{\rm i}$ is the sticking coefficient (in the range 0 to 1),
   $T_{\rm gas}$ is the gas kinetic temperature,
   $m_{\rm i}$ is the molecular mass of species $i$ in $amu$,
and $C$ is a factor which takes into account electrostatic effects (and is a function 
of the grain size, as defined in \citet{RHMW}). For neutral species, such as H$_2$O 
and CO, $C=1$.
$\sigma_{\rm H}$ is the grain population-averaged value of 
\[ 4\pi \overline{a}^2.d_{\rm g} \]
where $\overline{a}$ is the mean grain radius and 
$d_{\rm g}$ is the dust to gas ratio (by number, relative to $n_{\rm H}$) and
is effectively the mean `surface area of dust grains per hydrogen nucleon'.
The factor of 4 in this expression implicitly assumes that the grains are
spherical.

$\overline{a}$, and $\sigma_{\rm H}$ are important independent parameters in this 
study and in the models described below.
The significance of $a$, particularly in the determination of $\sigma_{\rm H}$, 
becomes particularly apparent when the effects of ice mantle growth are taken
into effect; thus, although the values of $A_{\rm th}$ and the number of accreted ice 
monolayers are not expected to be sensitive to $\overline{a}$, 
smaller values of $\overline{a}$ will result in a more rapid increase in $\overline{a}$
and $\sigma_{\rm H}$ as ice mantles grow; which can be termed  the `snowball effect'
So, for example, the growth of 50 monolayers of ice each of the same thickness) would 
typically result in an increase of the surface area by factors of $\sim 22$ and 
$\sim 1.4$ for grains whose bare radius is 0.01$\mu$m and 0.1$\mu$m respectively.

If we adopt the standard power-law grain size distribution for interstellar dust 
\citep{MRN} then $\overline{a}\sim 0.008\mu$m, the RMS value is 
$\overline{a^2}^{1/2}\sim 0.01\mu$m 
(appropriate for determining $\sigma_{\rm H}$), the median value of the 
grain surface area distribution is $\sim 0.018\mu$m and 
$\sigma_{\rm H} \approx 3.3\times 10^{-21}$cm$^2$. 
These values are highly uncertain, due to the assumptions about grain 
composition, morphology and the effects of grain aggregation and/or thick 
ice mantle accretion, but in any case it is clear that most of the surface area of the 
population resides in the smallest grains.

This value of $\overline{a}$ is significantly smaller than the value of $\sim 0.1\mu$m 
that is adopted in many studies \citep[e.g.][]{TCK12,Kal15}. Those studies often cite 
the likelihood of dust grain
agglomeration in dense environments but, if such processes are relevant, they will
also have very strong  - and usually neglected - effects on both $\sigma_{\rm H}$ and
the UV opacity and hence the $A_{\rm V}$-dependence of the photoreaction rates.

In the context of our discussion, it is also useful to define $f_{\rm sites}$, the
effective fractional abundance (relative to hydrogen nucleons) of binding sites:
\[ f_{\rm sites}= \sigma_H.N_{\rm s}  \]
where $N_{\rm s}$ is the mean surface density of binding sites on grains.
Thus, the surface coverage of a monolayer is given by the sum of the fractional 
abundances of solid-state species divided by $f_{\rm sites}$.

In the absence of any desorption processes, we can calculate 
(from equation~(\ref{eq:freeze}), above) a characteristic freeze-out timescale for 
oxygen atoms (for which $\mu$=18); $\tau_{\rm fo}$. $\tau_{\rm fo}$ is defined 
as the time required for the change in the abundance of gas-phase oxygen atoms 
to be comparable to the initial abundance. Thus
%
\begin{equation}
\tau_{\rm fo}=1.1Myr\left( \frac{10^{-20}{\rm cm}^{-2}}{\sigma_{\rm H}}\right)
\left( \frac{10}{T_{\rm gas}}\right)^{1/2}
\left( \frac{10^3 {\rm cm}^{-3}}{n_{\rm H}}\right) {\rm years} 
\label{eq:tfo}
\end{equation}
Note the somewhat counter-intuitive result that (assuming the 
dust grains are sufficiently cold for thermal desorption not to
occur), higher gas temperatures result in faster freeze-out. 

In reality, the `effective' sticking coefficient must take account of:
\begin{itemize}
\item[(a)] The sticking efficiency of O atoms, OH radicals and H$_2$O 
(and O$_2$) molecules,
\item[(b)] Surface reaction efficiencies for the conversion of O and OH to 
H$_2$O, and
\item[(c)] Desorption driven by the enthalpy of surface reactions. 
\end{itemize}

For (a) it is probably fair to assume that the sticking probabilities are of order unity.
For (b), reactions involving the hydrogenation of O, OH (and O$_2$) usually lead to 
H$_2$O formation. We recognise that the efficiencies are poorly constrained, and so we 
investigate a range formation efficiencies of 0-100\%, whilst 
for (c) we are guided by the empirical and theoretical studies of \citet{Min16}. The 
enthalpy of formation may result in the desorption of the product species and the 
chemical desorption efficiencies for the O+H$\to$OH reaction were determined to be 
$\sim 50$\% on oxidised graphite, and $\sim $25\% on amorphous water ice. For the 
OH+H$\to$H$_2$O reaction the desorption efficiencies are $\sim$50\%, $\sim$80\% and 
$\sim$30\% for oxidised grapite, amorphous silicates and amorphous water ice 
respectively. These imply a reduced net sticking/reaction efficiency for oxygen 
atoms and OH radicals.  

\subsection{Thermal desorption}
\label{subsec:thermal}

The rate of thermal sublimation (per molecule) for a zeroth order process is given by:
\[ k = \nu_0 e^{-E_{\rm bind}/k_{\rm B}T_{\rm dust}} {\rm ~s}^{-1} \]
where $E_{\rm bind}$ is the binding energy of the adsorbate, $T_{\rm dust}$ is the 
dust temperature and $\nu_0$ is the vibration frequency of the adsorbed molecule.
%
%
Due to the exponential dependence on the dust temperature, thermal desorption 
effectively operates as an on/off switch at a characteristic temperature defined by 
the balance between freeze out and thermal desorption. For H$_2$O if we balance 
the desorption rate with the freeze-out rate for oxygen atoms (assuming an effective 
sticking/hydrogenation coefficient of 1, a gas temperature 
of $\sim$10K and values for $N_{\rm s}$ and $E_{\rm bind}$ as given in 
Table~\ref{tab:param}), then the critical balance ($f_{\rm i}$=1) is obtained for
\begin{equation}
T_{\rm dust} \sim \frac{4800}{55-\ln n({\rm O})}
\label{eq:tdust}
\end{equation}
where $n(O)$ is the gas-phase density of oxygen atoms,
implying that thermal desorption is only efficient for $T_{\rm dust}>>10$K. It is 
therefore entirely suppressed for the conditions that exist within molecular clouds,
and is only a significant mechanism in the close vicinity of protostars, when the
requisite temperatures of $T_{\rm dust}\sim$100 K can be obtained.
By contrast, thermal desorption is much more important in molecular clouds for species 
such as CO, which are more volatile and have much lower surface binding energies.

\subsection{Cosmic ray induced desorption}
\label{subsec:crdes}

The rate of desorption of a species as a result of cosmic ray
heating of grains and ice mantles (whether in localised hot spots or whole grain 
heating) can be expressed as
\[ \dot{n}_{\rm i} = R_{\rm cr}n_{\rm i}  \]
where $R_{\rm cr}$ is the cosmic ray desorption rate per molecule and $n_{\rm i}$ is 
the volume abundance of the solid-state species \citep[e.g.][]{HH93}.
Importantly, the action of cosmic rays is not necessarily limited to the 
surface layers of the ices, so - unlike photodesorption - this rate depends on the 
total abundance of species in the ice and not just the surface coverage.
For the purpose of our numerical calculations note that,
in the critical case of $f_{\rm H_2O}=1$ (and, again, assuming a pure water ice), 
$n_{\rm H_2O}$ is given by $\sigma_{\rm H} N_{\rm s} n_{\rm H}$.

Until recently, calculations of the cosmic ray desorption rates have 
included a number of simplifications including the assumptions that (a) dust grains 
are spherical with high thermally conductivity, (b) the process is limited to larger 
(0.1$\mu$m), classical grains, (c) the various ice components co-desorb with 
characteristics defined for binding energies for simple, pure compositions, and 
(d) the cosmic-ray energy spectra are idealised and invariant. 
However, as is shown below, cosmic ray desorption is not likely to be a major 
contributor to the total H$_2$O desorption rate, so these issues are not relevant in 
this study.

The negligible importance of cosmic ray heating also seems to be supported by the 
findings of \citet{Wh13} who,
in their study of L183, found that the ices are amorphous in structure, indicating 
that they have not been heated to $\gtappeq$15\,K since their formation. 
This would seem to be consistent with the fact that - for this source at least - 
whole grain heating by cosmic rays is not an effective desorption mechanism,
unless the annealing effects of cosmic ray heating can be mitigated by exposure to
amorphizing UV radiation \citep[e.g.][]{LB03}.

\subsection{Photodesorption}
\label{subsec:photo}

Photodesorption from water ice initiated by the interstellar UV radiation field
provides an obvious explanation for the $A_{\rm th}$ phenomenon;
the photodesorption rate declines with depth into the cloud, so that at the cloud edge, 
photodesorption dominates over freeze-out and inhibits ice deposition, but at a 
sufficient depth, corresponding to $A_{\rm V}=A_{\rm th}$, freeze-out dominates over 
photodesorption and ice mantles begin to form. 

The photodesorption rate has been measured in the laboratory 
\citep[e.g.][]{Ob09,Yab06,West95} and has been the subject of many classical molecular 
dynamics simulations. The process involves the photodissociation of an H$_2$O molecule 
in the ice into energetic H and OH radicals which may recombine, escape, or be trapped 
in the ice lattice; the process depends on the depth within the ice of the dissociated 
molecule \citep{And08}.

The photodesorption flux (cm$^{-2}$) of a species $i$ is given by 
\citep[e.g.][]{HKBM09}:
\begin{equation}
F_{\rm pd,i} = Y_{\rm pd,i}F_{\rm UV}f_{\rm i} 
\label{eq:pdflux} 
\end{equation}
where $Y_{\rm pd,i}$ is the photodesorption yield for species $i$ and 
$F_{\rm UV}$ is the photon flux.
The rate of desorption of species $i$ is then given by:
\begin{equation}
\dot{n}_{\rm i} = \sigma_{\rm H} n_{\rm H} F_{\rm pd,i}
\label{eq:desi} 
\end{equation}
where $\sigma_{\rm H}. n_{\rm H}$ is effectively the grain surface area per unit volume
of the gas. The photon flux is given by:
\begin{equation}
F_{\rm UV} = GF_0e^{-1.8A_{\rm V}} + 4875.0\zeta
\label{eq:phflux} 
\end{equation}
In this expression the first term accounts for direct photodesorption; 
$F_0$ is the unshielded interstellar UV flux, with a
scaling factor $G$ and the exponential accounts for shielding by dust. 
The second term accounts for cosmic-ray induced photodesorption; $\zeta$ 
is a scaling factor for the cosmic ray ionization rate and is
normalised to the `standard' value of $\zeta_0=1.3\xten{-17}$s$^{-1}$.

$Y_{\rm pd,i}$ and $f_{\rm i}$ are both uncertain. $Y_{\rm pd,i}$ has been measured in 
the laboratory
to be $\sim 10^{-4}-10^{-3}$ but this is for pure water ices on a flat gold
substrate. It would seem unlikely that the same values would apply to 
impure/mixed ices on amorphous and morphologically complex grain surfaces.
It is also apparent that photodesorption is intrinsically a surface process which 
affects only the top few (maybe 3 or 4) monolayers of an ice mantle.
To calculate the photodesorption rate therefore requires a knowledge of the 
composition of the ices on a layer-by-layer basis.
The calculation of the fraction of surface binding sites that are occupied
by a species ($f_{\rm i}$) is non-trivial and requires a knowledge of the morphology 
and accretion history of individual ice mantles, as well as taking account of the 
adsorption
and desorption rates of all of the dominant ice constituents. Unless there
is complete fluid mixing within the ice $f_{\rm i}$ is {\em not} the same as the 
bulk mantle fractional composition of species $i$.

\subsection{Surface chemistry - chemically driven desorption}
\label{subsec:chemdes}

We have considered two mechanisms for desorption driven by the enthalpy of formation
of molecules in surface reactions:
\begin{itemize}
\item[a)] selective desorption of the product molecules (OH and H$_2$O) in surface 
hydrogenation reactions, and
\item[b)] non-selective desorption of molecules driven by the formation of H$_2$ 
on surfaces.
\end{itemize}
We have discussed the first of these in subsection~\ref{subsec:freeze} above, where we 
treat the process as effectively modifying the freezeout/reaction rate of O and OH.
For the the second, we know that the formation of H$_2$ molecules on the surface of 
dust grains results in an energy deposit into the grain that is a significant fraction 
of the H$_2$ bond energy (4.5eV).
Such a large energy input will be capable of desorbing molecules, possibly 
non-selectively, in the vicinity of an H$_2$ formation site, although the efficiency 
of that desorption is highly uncertain.

If we make the reasonable assumption that the desorption is a surface process (i.e.
effecting desorption from the surface layer(s) of the ice mantle) then 
we can write the rate of desorption (cm$^{-3}$s$^{-1}$) of a species $i$ as
\begin{equation}
\dot{n}_{\rm i} = Y_{\rm H_2,i}.\dot{n}_{\rm H_2}.f_{\rm i}
\label{eq:desrat} 
\end{equation}
where $Y_{\rm H_2,i}$ is the desorption yield for species $i$, $f_{\rm i}$ is the surface
coverage of species $i$, as above, and $\dot{n}_{\rm H_2}$ is the rate of (surface) 
H$_2$ formation.
Of course, $\dot{n}_{\rm H_2}$ is determined by the H-atom abundance which will, itself
be a function of $A_{\rm V}$.

For the purpose of this study, we can then speculate that the $A_{\rm V}$-dependence of 
ice formation may result from the variation in the abundances of O and H atoms with 
density and depth into a cloud.
As the H to H$_2$ abundance ratio declines with depth so too will the H$_2$-formation 
driven H$_2$O desorption rate until a point is reached (at $A_{\rm V}=A_{\rm th}$) at 
which it is less than the net accretion rate. 
If this were to prove a viable mechanism then, as the desorption yield is unknown, 
the observed value(s) of $A_{\rm th}$ could be used to predict the value(s) of
$Y_{\rm H_2,i}$.

Thus, balancing the expressions for freeze-out and chemical desorption rates given 
above, we see that (if chemical desorption is the main process that determines whether 
or not ices form) then the critical condition corresponding to $A_{\rm V}=A_{\rm th}$ 
is given by
\begin{equation}
\frac{n_{\rm O}}{n_{\rm H-atoms}} = 2 Y_{\rm H_2,H_2O} 
\left( \frac{S_{\rm eff,H}}{S_{\rm eff,O}} \right)
\label{eq:oatoms} 
\end{equation}
where $S_{\rm eff,O}$ and $S_{\rm eff,H}$ are the effective sticking and hydrogenation 
efficiencies for oxygen and hydrogen atoms, respectively. 
We assume these are of order unity and that the process is non-selective, to the extent
that it results in the desorption of strongly bound species, like H$_2$O 
\citep[an assumption that may not be valid e.g.][]{Rob07} and that the abundance of 
hydrogen atoms is $\sim 1$ cm$^{-3}$ in molecular clouds. 
With these assumptions, this implies that whilst H$_2$-formation driven 
desorption may be important in maintaining a significant gas-phase abundance of H$_2$O 
at high extinctions, it is only important in the determination of the abundance of 
solid-state H$_2$O when the density of oxygen atoms is $\ltappeq 0.5Y_{\rm H_2,H_2O}$.
With $Y_{\rm H_2,H_2O}\sim 10^{-4}-10^{-3}$ this condition is unlikely to be satisfied 
in normal circumstances.

\section{Analytical considerations}
\label{sec:analytical}

Adopting the standard values for the parameters as specified in Table~\ref{tab:param}, 
we find that the rates per hydrogen nucleon for the various desorption processes 
for H$_2$O; thermal, cosmic-ray
heating, direct photodesorption, cosmic ray induced desorption and desorption 
driven by the enthalpy of H$_2$ formation are:
$\sim$0, 1.45$\times 10^{-22}$, 3.3$\times 10^{-16}$e$^{-1.8A_{\rm V}}$, 1.6$\times 
10^{-20}$, and 1.9$\times 10^{-17}.Y_{\rm H_2}$
respectively. For comparison, the representative freeze-out rate is
$9.5\times 10^{-18} n_{\rm O}$.
From these values we can see that, at low extinctions, direct photodesorption 
is by far the dominant desorption mechanism.

However, as previously noted by \citet{WHW92} this result does not hold at higher 
extinctions. For example, with $A_{\rm V}=10$, the extinction in the UV absorption 
band for H$_2$O dissociation ($h\nu\sim$ 7 - 10 eV) is $\sim$ 30 magnitudes for a 
conventional interstellar extinction curve, so that external radiation fields are 
effectively excluded. In that case, photodesorption driven by the 
($A_{\rm V}$-independent) cosmic ray-induced UV radiation field \citep{PT83} becomes 
the dominant photodesorption mechanism, at almost unlimited depths in clouds or star 
forming regions.

From equation~\ref{eq:phflux} in subsection~\ref{subsec:photo} above we can quantify 
this;
cosmic-ray induced photodesorption dominates (and the desorption rate effectively 
ceases to have an explicit dependence on $A_{\rm V}$) once
\begin{equation}
A_{\rm V} > 5.51 + 0.56 \ln \left( \frac{G}{\zeta} \right)
\label{eq:avcr}
\end{equation}
Thus, whilst the $A_{\rm V}$-dependence can be invoked to explain $A_{\rm th}$ in Taurus
and Serpens, it cannot for $\rho$-Oph (where it would imply a value of 
$G/\zeta>10^5$).

Balancing the rate of freeze-out of oxygen atoms with the rate of photodesorption 
of H$_2$O molecules
(and neglecting the term due to cosmic-ray induced photodesorption) we obtain, with the 
value of $F_0$ given in Table~\ref{tab:param} and for $f_{\rm i}=1$: 
\begin{equation}
A_{\rm V} = 1.97 + 0.56\ln\left( G\frac{Y_{\rm pd}}{10^{-3}}\right) 
-0.56\ln\left( S_{\rm eff,O}.n_{\rm O} \right)
-0.28\ln\left( \frac{T_{\rm gas}}{10} \right) 
\label{eq:avpdes}
\end{equation}
From this expression we can see that, in regions where photodesorption is the dominant
desorption process, $A_{\rm th}$ is primarily dependent on the values of $G$, 
$Y_{\rm pd}$, $S_{\rm eff,O}$ and $n_{\rm O}$ (and hence the C$:$O ratio and the C to CO 
conversion efficiency - see below).
The dependence on $T_{\rm gas}$ is less significant as the range of uncertainty in the 
parameter is smaller than for those listed above.

If we adopt $G=1$, $Y_{\rm pd}=10^{-3}$ (as in Table~\ref{tab:param}), $S_{\rm eff,O}=1$ 
and $T_{\rm gas}=10$ then this formula implies that, to obtain $A_{\rm th}>0$, the 
oxygen atom abundance $n_{\rm O}\ltappeq 34$cm$^{-3}$. As the fractional abundance of 
free oxygen is of the order of $\sim 2\times 10^{-4}$ this implies that photodesorption 
cannot inhibit the formation of ices at any extinction if 
$n_{\rm H}\gtappeq 10^5$ cm$^{-3}$.

In regions where other desorption processes are significant, then the values of $\zeta$ 
and $Y_{\rm H_2}$ will also be important. However,
despite the wide variety of processes, there are not many other free parameters to which 
the condition for balance and $A_{\rm th}$ are sensitive. For example, the freeze-out 
rate and all of the desorption processes listed above are all proportional to the dust 
surface area ($\sigma_{\rm H}$) so that its value is not of critical importance.


\section{The chemical model}
\label{sec:model}

Whilst the above arguments are useful and yield qualititative and approximate 
quantitative results there are a number of inherent simplifications and approximations. 
For example, the abundance of oxygen atoms in equation~(\ref{eq:avpdes}) is implicitly 
dependent on the physical conditions, including the extinction.
To unravel this degeneracy a detailed chemical model has been 
applied to determine $A_{\rm th}$ in a self-consistent fashion.

The values of $A_{\rm th}$ and the slope of the water ice abundance vs. $A_{\rm V}$ 
($\alpha_{\rm H_2O}$) are near-universal in a specific cloud complex, indicating that the
material along the lines of sight probed by the observations has reached both gas-phase
and gas-grain chemical equilibrium. As such, time-dependent effects are not important in 
the determination of these parameters.
We have therefore investigated the sensitivities of $A_{\rm th}$ and $\alpha_{\rm H_2O}$
to the physical parameters, in a model of a dynamically static cloud, evolved to 
chemical equilibrium (at $t\sim 1-2\times 10^6$ years), although we have also 
considered the case of younger chemistries in test calculations.

The model is based on the comprehensive STARCHEM code that has been developed to study 
the chemistry in dynamically evolving star-forming regions, which has 
been successfully applied to the analysis of the chemistry of small molecular
species in the pre-stellar core L1544 \citep{RKC21}.

STARCHEM is a multi-point, flexible, model that can be adapted to a wide range 
of physical and chemical descriptions. The dynamical context can either be defined 
analytically, or by using an interpolated grid of physical parameters in arbitrary 
(1D or 2D) geometry.
In the current application, we simply assume a static, plane parallel, case. 
Abundances are then determined for each depth point and the integrated, line of sight, 
column densities of H$_2$O determined, for comparison with observations.
In STARCHEM, the chemistry is dynamically switchable and includes a full description
of the gas-phase, gas-grain and surface chemistry. For the purpose of this study 
we follow the depth and time-dependent evolution of the abundance of some 98 
gas-phase and solid-state species, involving the elements H, He, C, N, O, S and a 
representative low ionization potential metal; Na/Mg
in a network of 1239 reactions (1129 gas-phase, and 110 gas-grain reactions) that 
gives a good description of the chemistry of small species such as H$_2$O, CO etc., 
i.e. excluding large complex organic molecules. The chemistry is taken from 
the UDFA dataset \citep{udfa12}. Isothermal conditions are assumed. Although the 
densities are typically somewhat less than that required for the gas and dust 
temperatures to be well-coupled we have used a single temperature for both, noting 
that, for H$_2$O and in the range of parameters that we are exploring, the results 
are insensitive to the value of $T_{\rm dust}$.
Although STARCHEM includes an integrated model of photon-dominated regions (PDRs)
the values of $A_{\rm th}$ are $>2-3$ and so we can assume that the conditions 
are as appropriate to matter deep inside a PDR. We therefore assume that 
the photoionization rate of C and the photodissociation rates for H$_2$ and CO are 
negligible.

The chemistry is believed to be well-evolved, as argued above, so we do not 
follow the time-dependence of the (poorly constrained) surface chemistry. Rather,
we simply assume that most of the accreting O, O$^+$, O$_2$ and OH are converted to
H$_2$O (subject to some desorption due to the enthalpies of formation - as described 
below). 
It is assumed that the sticking efficiency of all species is 1.0. In addition,
we make some allowance for the partial ($\sim 10$\%) conversion of surface OH, in 
reaction with surface CO to form CO$_2$ to match the empirically determined 
CO$_2$:H$_2$O ratios in interstellar ices.
So, for example, accreting oxygen atoms are converted to OH radicals, a fraction
of which are returned to the gas-phase. A fraction of the remainder can react with 
surface CO to form surface CO$_2$, the rest is converted to H$_2$O and, again, a 
fraction of this is returned to the gas-phase, leaving the remainder as H$_2$O ice.

In our model we follow the procedures described in the STARCHEM model \citep{RKC21} 
and pay careful attention to (a) the composition of individual ice layers, and 
(b) the geometrical growth of grains due to ice deposition.
For (a) we note that an important consequence of the ice layering/surface layer 
desorption scenario is that the ices act as a `stack' that records and retains the 
prior chemical
evolution of a cloud; a fact that has been used, for example, to explain the abundance 
anomalies in cometary ices \citep{RWW19}.
Most desorption processes only occur in the upper layer(s) of the ices, so we 
model the layering of the ices and follow the chemical composition of each individual 
layer.
We then consider the two extreme examples of (i) individual layers retaining their 
chemical integrity (although subject to solid-state reactions taking place) - this is 
our standard assumption, and (ii) fully mixed ices, as would be more appropriate for
non-idealised ice layering and/or disruption by heating by, for example, cosmic rays. 
There is no allowance for surface porosity and,
other than the limited representation of the surface chemistry as described above, 
the ices themselves are treated as being chemically inert; that is to say we do not 
include the effects of chemical diffusion or mobility between layers or solid-state 
reactions within the ices. 
For (b) we make the assumption that grain growth and the increase in surface area 
occurs by spherically symmetric and layered accretion.

The chemical initial conditions are important and we assume that the gas starts from 
an essentially atomic state, but with significant conversion of H to H$_2$
($\sim$ 50\%) and C to CO (which we treat as a free parameter). In our standard model 
we assume that $\sim 90$\% of the carbon has been converted to CO, yielding an 
initial fractional abundance of CO of $2.7\times 10^{-4}$.

A full description and accurate quantification of the various desorption mechanisms 
as described above is included in the model.
In the case of H$_2$O photodesorption we follow \citet{HKBM09} and assume that - to 
model the observed OH$:$H$_2$O ratio in translucent clouds - the total dissociative 
desorption to OH + H and O + H$_2$ is twice as likely as non-dissociative 
photodesorption, and the total yield for the process is split accordingly.

The effect of desorption due to the enthalpy of formation of surface species is 
included for H$_2$ formation as an effective desorption rate with a generic 
(i.e. non-selective) yield for all species (as given in Table~\ref{tab:param}) and 
for OH and H$_2$O formation, as a modification of the effective sticking efficiency 
(otherwise taken to be unity) for the O+H and
OH+H surface reactions. The values used were taken from \citet{Min16} and allow for 
the differences between reactions occuring on bare grains ($f<1$) and icy 
mantles ($f>1$).  
We do not make any other allowance for variations in $S_{\rm eff}$ with surface 
composition in our standard model. 
The H$_2$O binding energy was taken to be 4800K \citep{DCN13}.

Other parameters are as specified in Table~\ref{tab:param}.
Note that the adopted values of the gas and dust temperatures have a marginal effect 
on the gas-phase chemistry and a weak effect (through a $T_{\rm gas}^{1/2}$ dependence)
on the freeze-out rates. The freeze-out rate of hydrogen atoms is used to evaluate
the H$_2$ formation rate self-consistently.
We adopt a representative gas density of $n=1.7\times 10^3$cm$^{-3}$ in our 
standard model.
This has been chosen to give the best fit and is somewhat arbitrary: a lower 
value of $n$ would result in a larger value of $A_{\rm th}$ 
(e.g. see Fig.~\ref{fig:results_fig2} ), but the range of density is is probably 
quite limited (see the discussion in section~\ref{sec:results}).

The elemental abundances have been taken from Table~5 (proto-solar abundances) from
\citet{AGSS09} although the oxygen abundance has been depleted 
from 5.4$\times 10^{-4}$ to a value of 5.0$\times 10^{-4}$ 
(as is consistent with our understanding of the oxygen budget, 
e.g. \citet{Wh10}). This is also chosen so as to obtain 
the best fit to the modelled $A_{\rm th}$-$N$(H$_2$O) correlation, as shown in 
Figure~\ref{fig:results_fig1}, and discussed below.

\begin{table*}
\centering
\caption{Standard parameters in the model.}
\label{tab:param}
\begin{tabular}{|l|l|l|}
\hline
 Parameter & Description & value \\
\hline
 $C/H$ & Total carbon abundance & 3.0$\xten{-4}$ \\
 $O/H$ & Total oxygen abundance & 5.0$\xten{-4}$ \\
 $X(CO)_0$ & Initial fractional abundance of CO & 2.7$\xten{-4}$ \\
 $T_{\rm gas}$ & Gas temperature & 10K \\
 $T_{\rm dust}$ & Dust temperature & 10K \\
 $\omega$ & Grain albedo &  0.5 \\
 $\overline{a}$ & Average grain radius & 8.3$\xten{-3}$ $\mu$m \\
 $\sigma_{\rm H}$ & Dust surface area per H nucleon & 3.3$\xten{-21}$ cm$^2$ \\
 $\Delta a$ & Ice monolayer thickness & 3.7\AA \\
 $S_{\rm eff,O}$ & Effective sticking and conversion coefficient (O$\to$H$_2$O) & 1.0 \\
 $F_0$ & Unshielded IS photon flux & 1.0$\xten{8}$ photons cm$^{-2}$ s$^{-1}$ \\
 $F_{\rm CR}$ & Cosmic ray-induced photon flux & 4875.0 photons cm$^{-2}$ s$^{-1}$ \\
 $\zeta$ & Cosmic ray ionization rate/1.3$\xten{-17}$ s$^{-1}$ & 1.0 \\
 $R_{\rm CR}$ & Cosmic ray desorption rate for H$_2$O & 4.40$\times 10^{-17}$s$^{-1}$ \\
 $Y_{\rm pd,i}$ & Photodesorption yield & 1.0$\xten{-3}$ \\
 $Y_{\rm pd-cr,i}$ & Cosmic ray induced photodesorption yield & 1.0$\xten{-3}$ \\
 $Y_{\rm H_2,i}$ & Yield for desorption by H$_2$ formation & 0.0 \\
 $G$ & ISRF/Draine  & 1.0 \\
 $N_{\rm s}$ & Grain binding site surface density & 1.0$\xten{15}$ cm$^{-2}$ \\
 $E_{\rm bind}/k$ & H$_2$O binding energy/k & 4,800\,K \\
\hline
\end{tabular}
\end{table*}

\section{Results and analysis}
\label{sec:results}

In the discussion that follows, we note that the essential constraints on the modelling 
are that the results must be consistent with (a) the observed values of 
$A_{\rm th}$, (b) the shape and slope of the $A_{\rm th}$-$N$(H$_2$O) correlation 
($\alpha_{\rm H_2O}$), and (c) the value of the `X-factor' (CO/H$_{\rm total}$).

We make the usual assumption that the total column density for hydrogen nucleons is 
related to $A_{\rm V}$ by $N$(H/cm$^{-2}$) $= 1.9\times 10^{21}$ $A_{\rm V}$ 
\citep[e.g.][]{Wh10}.
If we consider the observed correlation between $A_{\rm V}$ and $N_{\rm H_2O}$ given 
in equation~(\ref{eq:column}) then the implied line of sight averaged fractional 
abundance of solid-state H$_2$O is given by
\begin{equation}
X(H_2O)_{\rm s.} \sim 1.08\xten{-4}\left( 1 - \frac{2.7}{A_{\rm V}} \right), 
~A_{\rm V}>2.7
\label{eq:xh2o}
\end{equation}
indicating that, once the threshold for ice formation has been achieved, efficient 
freeze-out follows (the `snowball' effect).
Thus, $\alpha_{\rm H_2O}$ is a measure of the fractional abundance of H$_2$O in the 
solid state; X(H$_2$O)$_{s.}$.
This, in turn, depends on the amount of free oxygen in the (undepleted) gas-phase, 
most of which is subsequently converted to H$_2$O ice.
Unfortunately, this quantity is hard to estimate since (i) it will depend, in part, 
on the efficiency of conversion of C to CO, which is a relatively slow process, and 
(ii) there is a significant component of the interstellar oxygen budget that is 
unaccounted for. This component possibly accounts for in excess of 50\% of the total 
that is not included in the silicate/oxide dusts and may reside in some form of 
oxygen-bearing carbonaceous matter (Whittet, 2010).

Even allowing for this, at least 25\% of the remaining oxygen is taken up in the 
form of CO, if we assume that significant C to CO conversion has occured.
In dark clouds in the Milky Way it is usually reasonable to assume that this is a 
correct assumption, yielding a mean CO-to-H$_2$ conversion factor 
of X$_{\rm CO}\sim 2\xten{20}$cm$^{-2}$(K\,kms$^{-1}$). However, there is some
variation from sight line to sight line. In clouds with relatively low extinctions 
(i.e. $A_{\rm V}\sim 0.37-2.5$ magnitudes, so that they are at or near the 
threshold for CO detection, below which the emission region is classified as dark 
molecular gas) X$_{\rm CO}$ can be 6-7 times larger, suggesting that C to CO 
conversion is incomplete \citep{Luo20} in these sources. 
This may explain some of the scatter in the data points, particularly in the 
vicinity of $A_{\rm V}=A_{\rm th}$.

\subsection{$N$(H$_2$O) vs. $A_{\rm V}$}

\begin{figure*}
\includegraphics[width=1.8\columnwidth,trim=0 350 0 0]{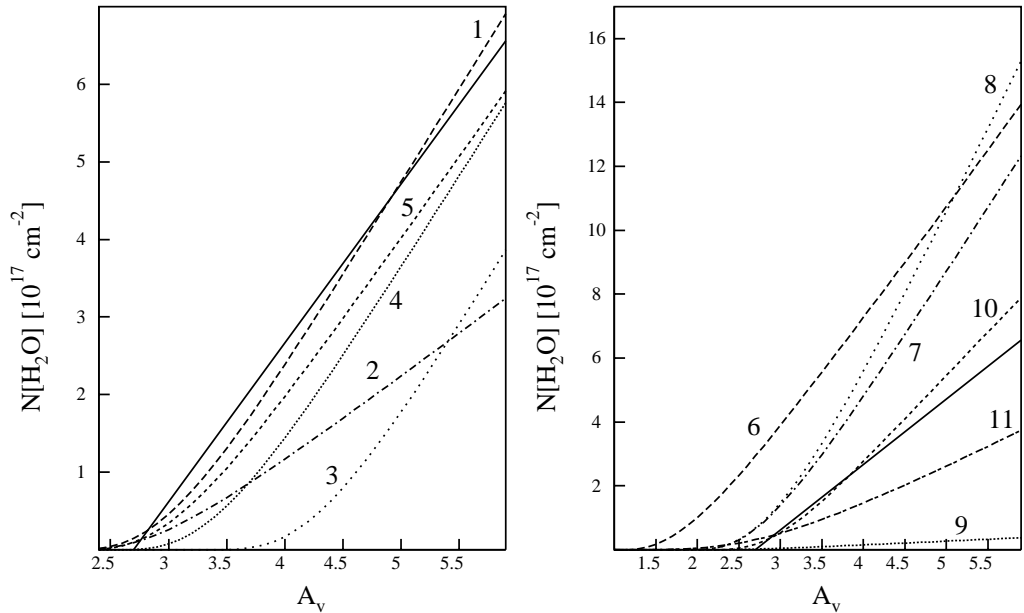}
\caption{The modelled late-time column densities of water ice as a function of 
extinction, $A_{\rm V}$. The labels refer to the model number, whose details are given 
in Table 2. The solid line in both plots is the empirical correlation (from data 
presented in \citet{Wh13}).}
\label{fig:results_fig1}
\end{figure*}

Results from the model are presented in two forms; as shown in 
figures~\ref{fig:results_fig1} and \ref{fig:results_fig2}.
In Figure~\ref{fig:results_fig1} we show the calculated late-time (equilibrium) 
values of the H$_2$O ice column density as a function of $A_{\rm V}$ (calculated in 
50 independent model calculations; 
spanning the range of extinctions $1<A_{\rm V}<5.9$) for variations in the free 
parameters, at a fixed density of 1,700 cm$^{-3}$. 

Although there are a large number of parameters in the model, only a relatively 
small subset of these have significant effects on $A_{\rm th}$ and/or the 
slope of the $A_{\rm V}$-$N$(H$_2$O) correlation for a given value of the density.
These are: $X({\rm O})_{\rm total}$ or $X({\rm CO})_0$, $G$, $Y_{\rm pd}$, $Y_{\rm H_2}$ 
(and the chemical desorption yields for OH and H$_2$O formation), $\sigma_{\rm H}$ and 
$\zeta$.
We consider the effects of variations in these parameters in models 1-11, whose
details are specified in Table~\ref{tab:models}.

\begin{table}
\centering
\caption{Parameter variations used in models 1-11.}
\label{tab:models}
\begin{tabular}{|l|l|l|}
\hline
 Model number & Parameter & value \\
\hline
 1 & standard & - \\
 2 & $\Delta a$ & 0 (no grain growth) \\
 3 & $G$ & 10.0 \\
 4 & $Y_{\rm pd}$/$Y_{\rm pd-cr}$ & 2.0$\xten{-3}$/3.0$\xten{-3}$ \\
 5 & $Y_{\rm H_2}$ & 2.0$\xten{-4}$ \\
 6 & $n$ & 1$\times 10^{4}$cm$^{-3}$ \\
 7 & - & no H$_2$O formation desorption \\
 8 & $X(CO)_0$ & $1.0\times 10^{-6}$ \\
 9 & Age & $10^5$ years \\
 10 & $\sigma_{\rm H}$ & 5$\xten{-21}$ cm$^2$ \\
 11 & O$\to$H$_2$O efficiency & 0.5 \\
\hline
\end{tabular}
\end{table}

As can be seen from Figure~\ref{fig:results_fig1}, the standard model (no. 1) gives a 
a good fit to $A_{\rm th}$, $\alpha_{\rm H_2O}$ and the shape of the 
$N$(H$_2$O)-$A_{\rm V}$ correlation although, as explained above, the initial value 
of the oxygen atom abundance was chosen to obtain the best fit to $\alpha_{\rm H_2O}$.

If we do not take into account the geometrical effects of ice mantle growth 
as would be appropriate for larger gain sizes, a very poor fit to
$\alpha_{\rm H_2O}$ is obtained (model 2) and cannot be rectified by changes in 
the other free parameters.
Increasing the radiation field strength (model 3) or using slightly higher 
photodesorption yields (model 4; for which $Y_{\rm pd}=2\xten{-3}$ and 
$Y_{\rm pd-cr}=3\xten{-3}$ for the direct and cosmic-ray induced processes 
respectively, as used in \citet{Kal15}) both strongly affect $A_{\rm th}$, 
although the effects on $\alpha_{\rm H_2O}$ are quite small.
Introducing H$_2$ formation-driven desorption with even a very small 
yield (model 5) also has a clearly discernable effect on $A_{\rm th}$.
Increasing the density (model 6) results in a significant reduction in $A_{\rm th}$, 
although the effect on $\alpha_{\rm H_2O}$ is relatively small.
Conversely, the non-inclusion of H$_2$O formation-driven desorption (model 7) or 
variations in the initial oxgen atom abundance (as controlled by the initial 
abundance of CO; model 8) both have strong effects on $\alpha_{\rm H_2O}$, although 
less so on $A_{\rm th}$.
Reducing the chemical age to 10$^5$ years dramatically inhibits ice formation at all 
extinctions (model 9).
As expected, the results are not strongly sensitive to the value 
of $\sigma_{\rm H}$ (model 10), but reducing the efficiency of the (solid state) 
conversion of O to H$_2$O results in significant reductions of both $A_{\rm th}$ and 
$\alpha_{\rm H_2O}$ (model 11).

In additional calculations (not shown in Figure~\ref{fig:results_fig1}) we found that 
increasing the cosmic ray ionization rate by a factor of 5-10$\times$ had very little 
effect on the results.
Also, if we arbitrarily reduce the sticking efficiency for oxygen (and other species)
from 1.0 to 0.5, then $A_{\rm th}$ is raised by $\sim$0.5 and $\alpha_{\rm H_2O}$ is 
significantly reduced.

%
%
%

\subsection{The dependence of $A_{\rm th}$ on the free parameters}

For Figure~\ref{fig:results_fig2} we have performed a large grid of calculations in 
($A_{\rm V}$, $n_{\rm tot.}$)-space, covering the range $1<A_{\rm V}<10.5$ and 
$2< log_{10} n <4.5$, again evolved until $t=1-2 \times 10^6$ years.
Calculations have been made for different values of the free parameters to
which $A_{\rm th}$ is most sensitive. 
The curves in the figures show the locus of points on the ($A_{\rm V}$, $n_{\rm tot.}$) 
plane corresponding to the conditions yielding $f_{\rm H_2O}=1$ which, as argued above, 
approximately corresponds to $A_{\rm V}=A_{\rm th}$.
This figure therefore shows the dependence of the modelled values of $A_{\rm th}$ on 
the local density ($n$), and how that dependence is sensitive to variations in other 
key free parameters.
The parameters/ranges that we have investigated and presented in the figure are:
(a) Radiation field strength: $G$= 0.5-20,
(b) Photodesorption yields: $Y_{\rm pd} = 10^{-4}$-$10^{-2}$,
(c) Cosmic ray ionization rate: $\zeta = 10^{-18}$s$^{-1}$-$10^{-16}$s$^{-1}$, and
(d) H$_2$-formation driven desorption yield: $Y_{\rm H_2} = 0-10^{-3}$.

Firstly, in all of the plots we note the sensitivity of $A_{\rm th}$ to the density 
($n$);
an increase in density of a factor of $\sim$10 results in a reduction of
$A_{\rm th}$ by $\sim$1.5 magnitudes.
Secondly, in most cases, no stable solution can be found for 
$A_{\rm th}\gtappeq 6-6.5$.
As expected, the correlations are seen to be sensitive to the values of $G$ and 
$Y_{\rm pd}$ although, to obtain a value of $A_{\rm th}$, as is appropriate for 
Serpens a very low density $\ltappeq 150$ cm$^{-3}$ is required.
The dependencies on $\zeta$ and $Y_{\rm H_2}$ are very much weaker. Large values 
of $Y_{\rm H_2}$ ($\gtappeq 5\times 10^{-4}$) result in significant modifications 
of the required minimum density for ices to form.

\subsection{The dependence on density}

The sensitivity of $A_{\rm th}$ to the density is hard to reconcile with the fact that 
$A_{\rm th}$ is very similar along different lines of sight unless the range of volume 
density is limited, and/or the volume density is closely correlated to the column 
density, which implies a self-similar evolution of the various cores that are probed 
along different lines of sight.

We have already noted (in section~\ref{sec:analytical}) that the balance between 
photodesorption and freeze-out requires that the density $<10^5$ cm$^{-3}$.
In addition, the uniformity of $\alpha_{\rm H_2O}$ implies that the freeze-out and 
desorption processes are in quasi-equilibrium. 
The timescale for freeze-out (given by equation~\ref{eq:tfo}) then implies that, for 
a reasonable upper limit of the cloud age of $\sim$10Myr, the density is 
$\gtappeq 3\times 10^2$ cm$^{-3}$ in the regions probed by the observations. 

\begin{figure*}
\includegraphics[width=1.8\columnwidth]{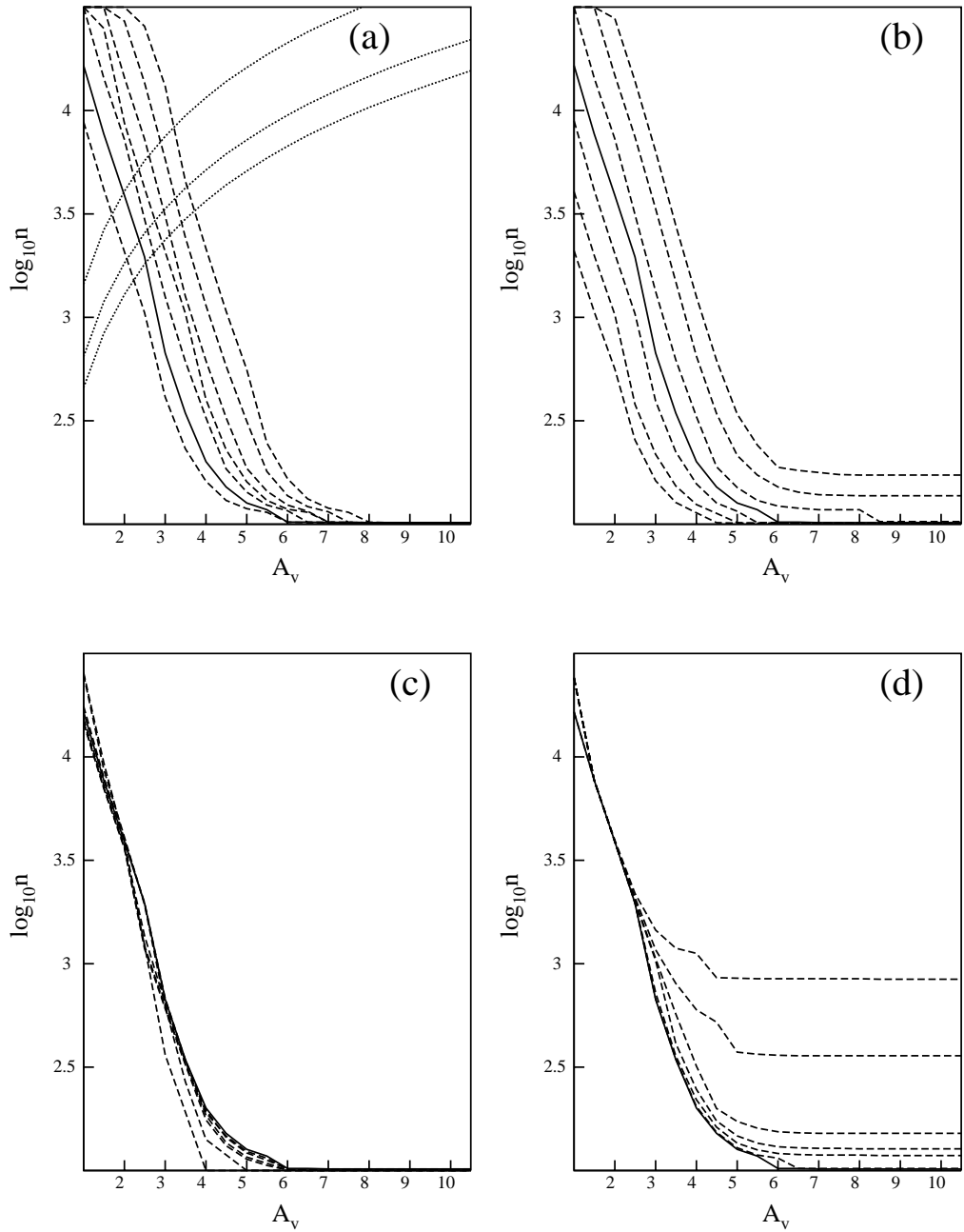}
\caption{The locus of points ($A_{\rm V}$,n$_{\rm tot.}$) for which 
$A_{\rm V}$=A$_{\rm th}$. 
The additional parameter that is varied in each plot, and the values for the curves 
(from bottom left to top right) are:
(a) $G$ = 0.5,1,2,3,5,10 and 20,
(b) $Y_{\rm pd}$ (in units of 10$^{-4}$) = 1,2,5,10,20,50 and 100,
(c) $\zeta=$ (in units of 10$^{-17}$s$^{-1}$) = 0.1,0.2,0.5,1.3,2,5 and 10,
(d) $Y_{\rm H_2}$ (in units of 10$^{-4}$) = 0,0.1,0.5,1,2,5 and 10.
The values corresponding to our standard model (1) are represented by the solid line 
in each plot.
The dotted lines in (a) show the correlation between density (n) and extinction 
($A_{\rm V}$) in the case of homologous collapse of uniform cores (from top to bottom) 
of masses 2, 10 and 20M$_\odot$.}
\label{fig:results_fig2}
\end{figure*}

This range of densities could be restricted further if we consider cloud structure 
and evolutionary state.
It is possible that the local density ($n_{\rm H}$) and $A_{\rm V}$ (and hence the 
column density $N_{\rm H}$) could be correlated if we assume that the lines
of sight pass through clumps/cores in varying states of evolution.
We can estimate this by assuming that each core is spherical, has a uniform density 
and contracts homologously (the approximate situation in the very earliest stages of 
collapse), in which case the column density and volume density of a core of mass $M$ 
are simply related by;
\begin{equation}
N = \left( \frac{6M}{\pi\mu m_{\rm H}} \right)^{1/3}n^{2/3}
\label{eq:colvol} 
\end{equation}
Assuming that the correlation between $N_{\rm H}$ and $A_{\rm V}$ holds this implies
that 
\begin{equation}
A_{\rm V} \sim 0.62 \left(\frac{M}{M_\odot}\right)^{1/3} 
\left(\frac{n}{1000~{\rm cm}^{-3}} \right)^{2/3} 
\label{eq:avm}
\end{equation}
We have plotted the correlation given by equation~(\ref{eq:avm}) in 
Fig.~\ref{fig:results_fig2} for the cases of $M$=2, 10 and 20M$_\odot$. 
What this shows is that, for our standard model - with this additional constraint 
for this range of core masses - the upper limit of the density is significantly 
reduced to 
$\sim 4\times 10^3$cm$^{-3}$ and the value of $A_{\rm th}$ lies
in the relatively narrow range of $\sim 2-2.7$ (and is perhaps indicative of a 
slightly higher value of $G\sim 2$).

If the cores are more evolved than this simple uniform density representation, then 
their internal structures will be closer to that of Bonnor-Ebert spheres, but a 
similar dependence would result, with a somewhat different scaling factor.
In any case, these results indicate that, although there are a large number of free 
parameters in the model, they are all fairly well-constrained to lie within 
relatively narrow ranges. 

As a caveat, we note that the situation could be further complicated if we recognise 
that $A_{\rm V}>A_{\rm th}$ in the innermost parts of cores, in which case line of 
sight filling factors must be taken into account.

\subsection{Summary of results}

In broad terms the resulting dependencies on the free parameters in the model are 
found to be as follows:
\begin{enumerate}
\item The value of $A_{\rm th}$ is, as expected, strongly sensitive to (a) the 
radiation field strength ($G$), (b) the yields for the photodesorption processes, 
and (c) the local density. 
\item $A_{\rm th}$ has a weaker dependence on the abundance of free oxygen, and the
assumed H$_2$O formation-driven desorption and surface hydrogenation efficiencies.
\item Other than variations in the instrinsic microphysics (e.g. the dust grain surface 
area per hydrogen nucleon, and whether or not ice mantle growth is taken into account, 
the efficiencies of H$_2$O formation-driven desorption and surface hydrogenation 
reactions), the slope of the $N$(H$_2$O)-$A_{\rm V}$ correlation ($\alpha_{\rm H_2O}$) 
is only sensitive to the initial abundance of free oxygen atoms and the assumption of
chemical equilibrium (i.e. the chemical age).
\item The assumption of no grain growth due to ice mantle accretion (which is 
close to being accurate in the case of an assumed larger initial mean grain radius) 
results in a large reduction of $\alpha_{\rm H_2O}$.
\end{enumerate} 

\section{Alternative mechanisms}
\label{sec:alt}

In this section we investigate two other possible mechanisms that may explain the 
$A_{\rm th}$ phenomenon, that are not directly related to the $A_{\rm V}$-dependence of
photodesorption. Both of these rely on variations in the binding energy of H$_2$O to 
dust grains.

The empirical basis for the determination of desorption rates are generally based
on laboratory experiments involving pure or mixed multi-layer ices that cover idealised
carbon (graphitic, metallic) substrates. However, if we consider both (a) the low 
surface coverage (sub-monolayer), or (b) different forms of the substrate, then 
significant variations in the net sticking efficiency for oxygen atoms may result.

\subsection{Threshold coverage for hydrogen bonding}
\label{subsec:hbond}

Firstly, we speculate that the onset of ice deposition at a particular visual extinction
is determined by the requirement that the flux of impinging atoms must exceed some 
critical value so as to maintain a significant surface coverage of ice. 
The assumption is that lone H$_2$O molecules are relatively weakly bound to the 
surface and can readily be desorbed by various mechanisms, whereas bulk water ice is 
bound by relatively much stronger hydrogen bonding, in addition. Thus, for ice to be 
formed, the O atom flux at the surface of grains must be large enough to overcome any 
desorbing mechanisms, by establishing a surface density of H$_2$O molecules above some 
critical value.

We can simulate this process, in effect, by enhancing the desorption rates if the 
abundance of H$_2$O in the sub-monolayer regime is less than some critical value,
$f_{\rm H_2O,crit}$.
To test this, we have performed calculations where we (arbitrarily) enhance all of the 
desorption rates by factors of (i) 10$\times$ and (ii) 100$\times$ in the sub-monolayer
regime, and take the extreme example of $f_{\rm H_2O,crit}=1$. In practice, this has to
be applied to all species, and not just H$_2$O, otherwise ices of CH$_4$, NH$_3$ and
CO etc. rapidly establish a monolayer and the effect is fairly minimal, even if the
desorption enhancement is applied to the top three monolayers.
Applying the enhancement to all species results in a significant increase in the values
of $A_{\rm th}$ (e.g. to 4.4 and 5.6 for models 1 and 3, respectively, for case(i)). 
For case (ii) ices never form at any value of $A_{\rm V}$ for models 1-5. However, in 
all models the situation rapidly switches once $f_{\rm H_2O,crit}$ has been obtained and 
the usual $N$(H$_2$O) - $A_{\rm V}$ correlation ensues. This yields a discontinuous 
curve that is at odds with the observations. We therefore conclude that this process is 
probably not important. 
  
\subsection{The dependence of H$_2$O bonding with grain surface on $A_{\rm V}$}
\label{subsec:carbon}

The nature of the surface of dust grains changes with depth into a cloud, and here we 
speculate that this may affect the binding energy of H$_2$O to the surface.
Interstellar dust is conventionally assumed to be composed predominantly of silicates 
or carbons. Experimental and theoretical studies of H$_2$O binding energies are mostly 
limited to carbons, in particular to graphitic (sp$^2$) carbon. This may not be too 
great a restriction, as in some models of interstellar dust, much of the carbon is 
assumed to be in the form of a layer on silicate substrates rather than free-flying 
carbon grains \citep[e.g.][]{Jones90}. If so, then the H$_2$O-silicate interaction 
is of lesser importance.

However, hydrogenated amorphous carbon may adopt several kinds of chemical bonding, 
including sp$^2$ (graphitic) and sp$^3$ (polymeric); the bonding affects the physical
and chemical properties of the solid. Conversions between these and other types of 
carbon may be induced by UV irradiation (photodarkening; sp$^3\to$sp$^2$) and by hot 
H atom (and C/C$^+$ atom) insertion (sp$^2\to$sp$^3$).
Thus, interstellar carbons may be expected to change their nature according to their 
environment \citep{Jones90}. Detailed studies \citep[e.g.][]{CCP10} show that for 
typical interstellar cloud conditions the bulk of the carbon on a silicate grain is 
sp$^2$; however, much of the carbon surface layer is sp$^3$.  For molecular clouds, 
after a reasonable evolution time ($\sim 10^5$ years) a phase transition between 
sp$^2$ and sp$^3$ occurs in the surface layer at depths into a cloud of about 3 
magnitudes. H$_2$O is likely to have a lower binding energy with sp$^2$ carbon than it 
does with hydrogenated sp$^3$ carbon, via hydrogen bonding.
It is therefore possible that this phase transition may affect the deposition of 
H$_2$O on interstellar dust and it is interesting to note that the transition occurs 
at depths similar to values of $A_{\rm th}$ in nearby interstellar clouds. 

Unfortunately, whilst the binding energy of H$_2$O on graphite and other sp$^2$ 
surfaces has been well determined and forms the basis of empirical studies of 
laboratory simulations of interstellar gas-grain interactions, the binding energy 
on polymeric carbon (sp$^3$) is currently unknown and so we are not able to quantify 
the effect.
However we note that, unlike the other mechanisms discussed in this paper, this 
provides a possible explanation for the $A_{\rm th}$ phenomenon that is unrelated to 
the $A_{\rm V}$-dependence of photodesorption.

%

\section{Conclusions}
\label{sec:conc}

Firstly, we should note that
there are several simplifications and assumptions that we have made in this paper:
\begin{enumerate}
\item Although we have considered grain growth by ice mantle accretion and its 
regulation of the freeze-out and desorption processes, we have not included the effect
that such grain growth will have on the UV extinction and how that would modify the
dependence of photoprocesses (including photodesorption) on $A_{\rm V}$.
Thus, for example, in our standard model once $A_{\rm V}\sim 4$, $\sim$18 ice layers
will have been accreted, implying an ice mantle thickness of $\sim 70$\AA. This 
effectively eliminates the population of particles with sizes 
50\AA\ $<a<$ 120\AA\ from the distribution and would, inevitably, result in a 
significant reduction of the extinction in the UV.  
%
%
\item We have not included the effects of changes in the grain
albedo as ices build up. 
There is indeed evidence for a change in the optical properies of the dust
as the density increases \citep{Wh13} and such changes could result in significant 
variations in the cosmic-ray induced photon flux and photodesorption rates.
\item We have treated clouds as plane parallel and have not included any representation 
of the internal density structure.
We could, for example, speculate an alternative theory for the origin of the 
$N$(H$_2$O)-$A_{\rm V}$ correlation that is based on the observational evidence that 
clouds contain transient microstructure \citep{Hetal03,Gir02,Garrod06}.
The densities in the microstructure appear to be high enough for freeze-out to be 
rapid. In this picture, a dark cloud may be considered as a collection of dense 
clumps embedded in a lower density background gas, some of which may be 
gravitationally bound and subsequently collapse to form stars. 
The ice would be entirely contained within 
these clumps, that may be intersected by a line of sight passing through the cloud. 
Thus, $N$(H$_2$O) would be determined by the clump filling factor in the 
observational beam and/or the number of clumps that intersect the line of sight, and 
$\alpha_{\rm H_2O}$ would be determined by the density and distribution of clumps.

\end{enumerate}

However, despite these assumptions and limitations, we can draw some important 
conclusions from this study. Our main findings are:
\begin{itemize}

\item The fact that $A_{\rm th} \sim$ 3.2 in several quiescent regions
(Taurus, Elias16, IC 5146) is remarkable and points to a common cause that is neither 
time-dependent, nor sensitive to local variations of physical conditions.

\item We have shown that the astronomical data for $A_{\rm th}$ and 
$\alpha_{\rm H_2O}$ are not only powerful probes of the physical
conditions within dark interstellar clouds but also constrain the
poorly-known microphysics of ice deposition in those regions.

\item The existence of $A_{\rm th}$ can be explained as being due to the balance 
between freeze-out and photodesorption for these sources. No other desorption 
mechanism (even the more speculative ones) is capable of explaining the phenomenon.

\item We have investigated two other mechanisms as possible causes of the $A_{\rm th}$ 
phenomenon, in which the effective binding energy of H$_2$O molecules is determined by
(i) the requirement of a critical surface density of H$_2$O, or
(ii) the nature of the surface layers of carbon (sp$^2$ or sp$^3$).
We find that the first of these can result in significant variations in $A_{\rm th}$, 
but is still subject to the same constraints as simple photodesorption, and predicts 
an incorrect shape to the $A_{\rm V}$-$N$(H$_2$O) correlation. The second is viable, 
and is largely independent of the photodesorption constraints, but is - as yet - 
unquantifiable. 

\item In addition to $A_{\rm th}$, the near-linear shape and slope 
($\alpha_{\rm H_2O}$) of the correlation between the extinction and the inferred 
H$_2$O ice column density is also very similar along different lines of sight. 
This implies that the observations probe clouds in remarkably similar dynamical 
and chemical evolutionary states.
Our model yields a very good match to the value of $A_{\rm th}$, $\alpha_{\rm H_2O}$ 
and the shape of the $A_{\rm V}$-$N$(H$_2$O) correlation. 
However, the values of $A_{\rm th}$ and $\alpha_{\rm H_2O}$ are affected by the 
values of the key free parameters that we have investigated in our models 
(most notably the density, the radiation field strength, the oxygen abundance 
and the characteristics of the dust grain physics and surface chemistry), implying 
that none of them can vary significantly along different lines of sight in the Taurus 
complex.

\item The near linear shape of the $A_{\rm V}$-$N$(H$_2$O) correlation is consistent 
with the fact that the bulk of the ice grows on (initially) very small dust grains 
($a<0.01\mu$m).

\item Once ice starts to be deposited, then it rapidly builds up to saturation levels, 
containing almost all of the available oxygen that is not bound in CO or organic 
or refractory components.
The shape and slope of the $A_{\rm V}$-$N$(H$_2$O) correlation indicates that
the clouds are well-evolved (with ages $>>10^5$ years) and with a uniform net 
free oxygen abundance (prior to freeze-out) of $\sim 2.3\times 10^{-4}$
(indicating efficient conversion of C to CO).

\item The only ways in which significantly larger values of $A_{\rm th}$ can be 
obtained are (i) by increasing the ambient radiation field strength ($G$), 
(ii) increasing the photodesorption yields ($Y_{\rm pd}$), and/or 
(iii) reducing the density ($n$).
The value of $A_{\rm th}$ is only sensitive to these parameters.  
The differences between $A_{\rm th}$ in Taurus and Serpens can probably best be 
explained by differences in $G$. Although less likely as the cause, the differences 
could possibly originate from extreme variations in the dust composition and 
hence the desorption yields. 

\item By contrast, the very high value of $A_{\rm th}\sim 10-15$ that is determined for 
$\rho$-Oph cannot be explained so simply; photodesorption by secondary (cosmic-ray
induced) photons would be the more significant mechanism at extinctions
significantly less that these values. 
The fact that there {\em is} a correlation between $A_{\rm V}$ and $N_{\rm H}$ 
indicates that some other process is important. $\rho$-Oph is more dynamically 
active than Taurus or Serpens, so that internal heating sources may be important. 

Alternatively, it is possible that UV photons could penetrate deep inside a cloud 
if the extinction is anomalous in the far-UV rise. In dense gas, the effects of both 
ice mantle accretion and dust coagulation become effective and the relative absence 
of small dust grains could greatly enhance the UV field strength at much higher 
$A_{\rm V}$ values than normal.
 
\item $A_{\rm th}$ has a fairly strong dependence on the gas density which implies
that the range of densities probed by observations is restricted and/or there is a 
correlation between volume and column densities.
The requirements of chemical equilibrium and the balance between photodesorption and 
freeze-out set some constraints on the density range, but this is significantly 
restricted if the assumption is made that the cores are in different stages of 
evolution along a similar evolutionary track.
The universality of $A_{\rm th}$ and $\alpha_{\rm H_2O}$ is therefore remarkable as it 
implies that, averaged over the various lines of sight, there must be near-universal 
laws of morphological/density structure and evolution within the cloud complexes.

\end{itemize}

\section*{Data availability}

The data underlying this study are openly available from the published
papers that are cited in the article.
No new data were generated in support of this research.

\bsp
\label{lastpage}

\begin{thebibliography}{}


\bibitem[Andersson \& van Dishoeck(2008)]{And08}
Andersson S., van Dishoeck E.F., 2008, A\&A, 491, 907

\bibitem[Asplund et al.(2009)]{AGSS09}
Asplund M., Grevesse N., Sauval A.J., Scott P., 2009,
Ann Rev A\&A, 47, 481

\bibitem[Bergin et al.(2005)]{Bergin05}
Bergin E., Melnick G.J., Gerakines P.A., Neufeld D.A., Whittet D.C.B, 
2005, ApJ 627, L33

\bibitem[Cecchi-Pestellini et al.(2010)]{CCP10}
Cecchi-Pestellini C., Cacciola A., Iati M.A., Saija R., Borghese F.,
Denti P., Giusto A., Williams D.A., 2010, MNRAS, 408, 535

\bibitem[Chiar et al.(1994)]{Chiar94}
Chiar J.E., Adamson A.J., Kerr T.H., Whittet D.C.B., 1994, ApJ, 426, 240

\bibitem[Chiar et al.(2007)]{Chiar07}
Chiar J.E., et al., 2007, ApJ, 666, L73

\bibitem[Chiar et al.(2011)]{Chiar11}
Chiar J.E., Pendleton Y.J, Allamandola L.J., et al., 2011,
ApJ, 731, 9

\bibitem[Cuppen \& Herbst(2007)]{CH07}
Cuppen H.M., Herbst E., 2007, ApJ, 668, 294

\bibitem[Cuppen et al.(2010)]{Cup10}
Cuppen H.M., Ioppolo S., Romanzin C., Linnartz H., 2010, Phys.Chem. Chem.
Phys., 12, 12077

\bibitem[Du et al.(2012)]{Du12}
Du F., Parise B., Bergman P., 2012, A\&A, 538, A91

\bibitem[Dulieu et al.(2013)]{DCN13}
Dulieu F., Congiu E., Noble J. et al., 2013, Nature Sci. Rep., 3, 1338

\bibitem[Eiroa \& Hodapp(1989)]{EH89}
Eiroa \& Hodapp, 1989, A\&A, 201, 354

\bibitem[Garrod et al.(2006)]{Garrod06}
Garrod R.T., Williams D.A., Rawlings J.M.C., 2006, ApJ, 638, 827

\bibitem[Gillett \& Forrest(1973)]{GF73}
Gillett F.C., Forrest W.J., 1973, ApJ, 179, 483

\bibitem[Girart et al.(2002)]{Gir02}
Girart J.M., Viti S., Williams D.A., Estalella R., Ho P.T.P., 
2002, A\&A, 388, 1004

\bibitem[Hartquist et al.(2003)]{Hetal03}
Hartquist T.W., Falle S.A.E.G., Williams D.A., 2003, ApSpSci, 288, 369

\bibitem[Hasegawa \& Herbst(1993)]{HH93}
Hasegawa T., Herbst E., 1993, MNRAS, 261, 83

\bibitem[Hollenbach et al.(2009)]{HKBM09}
Hollenbach D., Kaufman M.J., Bergin E.A., Melnick G.J., 2009,
ApJ, 690, 1497

\bibitem[Ioppolo et al.(2008)]{Ietal08}
Ioppolo S., Cuppen H.M., Romanzin C., 2008, ApJ, 686, 1474


\bibitem[Jones \& Williams(1984)]{JW84}
Jones A.P., Williams D.A., 1984, MNRAS, 209, 955

\bibitem[Jones et al.(1990)]{Jones90}
Jones A.P., Duley W.W., Williams D.A., 1990, QJRAS, 31, 567

\bibitem[Kalv\~{a}ns(2015)]{Kal15}
Kalv\~{a}ns J., 2015, ApJ, 803, 52

\bibitem[Lamberts et al.(2013)]{Lam13}
Lamberts T., Cuppen H.M., Ioppolo S., Linnartz H., 2013, 
Phys. Chem. Chem.Phys., 15, 8287

\bibitem[Leto \& Baratta(2003)]{LB03}
Leto G., Baratta G.A., 2003, A\&A, 397, 7

\bibitem[Linnartz(2011)]{Lin11}
Linnartz H., 2011, MNRAS, 413, 2281

\bibitem[Luo et al.(2020)]{Luo20}
Luo G. et al., 2020, ApJ, 889, L4

\bibitem[Mathis, Rumpl and Nordsieck(1977)]{MRN}
Mathis J.S., Rumpl W., Nordsieck K.H., 1977, ApJ, 217, 425

\bibitem[McElroy et al.(2013)]{udfa12} 
McElroy D., Walsh C., Markwick A.J.,
Cordiner M.A., Smith K., Millar T.J., 2013, A\&A, 550, A36

\bibitem[Minissale et al.(2016)]{Min16}
Minissale M., Dulieu F., Cazaux S., Hocuk S., 2016, A\&A, 585, A24

\bibitem[Noble et al.(2013)]{Noble13}
Noble J.A., Fraser H.J., Aikawa Y., Pontoppidan K.M., Sakon I., 2013,
ApJ, 775, 85

\bibitem[\"{O}berg et al.(2009)]{Ob09}
\"{O}berg K. I., Linnartz H., Visser R., van Dishoeck E.F., 
2009, ApJ, 693, 1209

\bibitem[Prasad \& Tarafdar(1983)]{PT83}
Prasad S.S., Tarafdar S.P., 1983, ApJ, 267, 603

\bibitem[Rawlings et al.(1992)]{RHMW}
Rawlings J.M.C., Hartquist T.W., Menten K., Williams D.A., 1992,
MNRAS, 255, 471


\bibitem[Rawlings, Wilson \& Williams(2019)]{RWW19}
Rawlings J.M.C, Wilson T.G., Williams D.A., 2019, MNRAS, 486, 10

\bibitem[Rawlings, Keto \& Caselli(2021)]{RKC21}
Rawlings J.M.C., Keto E., Caselli P., 2021, in preparation

\bibitem[Roberts et al.(2007)]{Rob07}
Roberts J.F., Rawlings J.M.C., Viti S., Williams D.A., 2007, MNRAS, 382, 733



\bibitem[Tanaka et al.(1990)]{Tan90}
Tanaka M., Sato S., Nagata T., Yamamoto T., 1990, ApJ, 352, 724

\bibitem[Taquet, Ceccarelli \& Kahane(2012)]{TCK12}
Taquet V., Ceccarelli C., Kahane C., 2012, A\&A, 538, A42

\bibitem[van Dishoeck et al.(2013)]{vD13}
van Dishoeck E.F., Herbst E., Neufeld D., 2013, Chem Rev, 113, 9043

\bibitem[Westley et al.(1995)]{West95}
Westley M.S., Baragiola R.A., Johnson R.E., Baratta G.A., 
1995, Nature, 373, 405

\bibitem[Whittet et al.(1989)]{Wh89}
Whittet D.C.B., Adamson A.J., Duley W.W., Geballe T.T., McFadzean A.D.,
1989, MNRAS, 241, 707

\bibitem[Whittet \& Blades(1980)]{WB80}
Whittet D.C.B., Blades J.C., 1980, MNRAS, 190, 403

\bibitem[Whittet et al.(2001)]{Wh01}
Whittet D.C.B., Gerakines P.A., Hough J.H., Shenoy S.S., 2001,
ApJ, 547, 872

\bibitem[Whittet(2003)]{Wh03}
Whittet D.C.B., 2003, Dust in the Galactic Environment, IoP Publishing,
Bristol

\bibitem[Whittet et al.(2007)]{Wh07}
Whittet D.C.B., Shenoy S.S., Bergin E.A., Chiar J.E., 
Gerakines P.A., Gibb E.L., Melnick G.J., Neufeld D.A., 2007,
ApJ, 655, 332

\bibitem[Whittet et al.(2009)]{Wh09}
Whittet D.C.B., Cook A.M., Chiar J.E., et al., 2009, ApJ, 695, 94

\bibitem[Whittet(2010)]{Wh10}
Whittet D.C.B., 2010, ApJ, 710, 1009

\bibitem[Whittet et al.(2013)]{Wh13}
Whittet D.C.B., Poteet C.A., Chiar J.E., Pagani L., Bajaj V.M., 
Horne D., Shenoy S.S., Adamson A.J., 2013, ApJ, 774, 102

\bibitem[Williams et al.(1992)]{WHW92}
Williams D.A., Hartquist T.W., Whittet D.C.B., 1992, MNRAS, 258 599

\bibitem[Yabushita et al.(2006)]{Yab06}
Yabushita A., Kanda D., Kawanaka N., Kawasaka M., Ashfold M.N.R.,
2006, J Chem Phys, 125, 3406

\end{thebibliography}
\end{document}